\documentclass[preprint,epsfig,onecolumn,floats,showpacs]{revtex4}
\usepackage{amssymb}
\usepackage{graphicx}
\usepackage{epsfig}
\usepackage{amsmath}
\usepackage{amsfonts}
\usepackage{bm}
\usepackage{subfigure}

\begin{document}

\title{Coulomb blockade in graphene quantum dots}
\author{Qiong Ma}
\author{Tao Tu}
\email{tutao@ustc.edu.cn}
\author{Zhi-Rong Lin}
\author{Guang-Can Guo}
\author{Guo-Ping Guo}
\email{gpguo@ustc.edu.cn}
\affiliation{Key Laboratory of Quantum Information, University of Science and Technology
of China, Chinese Academy of Sciences, Hefei, 230026, P.R.China}
\date{\today}

\begin{abstract}
We study the conductance spectrum of graphene quantum dots, both single and
multiple cases. The single electron tunneling phenomenon is investigated and
the periodicity, amplitude and line shape of the Coulomb blockade
oscillations at low temperatures are obtained. Further, we discuss the
transport behavior when multiple dots are assembled in array and find a
phase transition of conductance spectra from individual Coulomb blockade to
collective Coulomb blockade.
\end{abstract}

\maketitle

\baselineskip 16pt

\textit{Introduction.} After first isolation from bulk graphite, graphene
has attracted intense experimental and theoretical attention due to its
unusual electronic spectrum and hence exotic properties \cite%
{Geim2007,Geim2009}. Moreover, its zero nuclear spin and weak spin-orbit
interaction make it one of the most promising platforms for solid state
quantum information processing. Design and fabrication of graphene quantum
dot to confine Dirac fermion should be a critical step towards realizing
such promise. There are alternative theoretical schemes proposed to deal
with graphene's gapless spectrum and form quantum dot \cite%
{Efetov2007,Burkard2007,Peeter2007,Egger2007}. Recently, single or double
graphene quantum dots by etching nanoribbon into a charge island have been
demonstrated \cite{Geim2008,Ensslin2008-1,Ensslin2008-2,Ensslin2009-1}.
Although there are some initial studies \cite{Brouwer2009,Wurm2009}, the
transport properties of graphene quantum dot are still remain to be explored
both experimentally and theoretically.

In this paper, we study conductance of graphene quantum dots in both single
dot and multiple dots regime, respectively. In previous theoretical studies,
the conductance resonances are determined by the level of the (quasi-)bound
states of the noninteracting electrons \cite{Efetov2007,Brouwer2009,Wurm2009}%
. In the experiment the shape of the resonances will be governed by the
electron interaction via the Coulomb blockade (CB) effect \cite%
{Geim2008,Ensslin2008-1,Ensslin2008-2}. We use the Constant Interaction (CI)
model to account for the intradot Coulomb interaction and perform an exact
numerical calculation of the low bias conductance within quantum transport
formalism \cite{Beenakker1990}. The obtained CB peaks provides a simple way
to describe the rich phenomena of transport physics in a variety of graphene
quantum dots. In particular, we find the interdot tunneling would lead to a
phase transition from individual CB to collective CB in the conductance
spectra through multiple graphene quantum dots.

\textit{Graphene quantum dot. }We begin by considering the graphene quantum
dot on electrostatically confined graphene nanoribbon with semiconducting
armchair boundary \cite{Burkard2007}. The electron wave function in graphene
is usually described as a four-component (iso)spinor $\Psi =(\Psi
_{A}^{(K)},\Psi _{B}^{(K)},\Psi _{A}^{(K^{^{\prime }})},\Psi
_{B}^{(K^{^{\prime }})})$. Close to the Dirac point, i.e. on low energy
scales, a weakly doped (or undoped) graphene can be described by two
identical and decoupled Dirac Hamiltonian
\begin{eqnarray}
H_{K} &=&v_{F}\sigma \cdot p+U(x,y), \\
H_{K^{^{\prime }}} &=&v_{F}\sigma ^{\ast }\cdot p+U(x,y),
\end{eqnarray}%
where the Fermi velocity $v_{F}\approx 10^{6}$ m/s, the momentum operator $%
p=-i\hbar (\partial _{x},\partial _{y})^{T}$, Pauli matrices $\sigma
=(\sigma _{x}$, $\sigma _{y})$, $\sigma ^{\ast }=(\sigma _{x}$, $-\sigma
_{y})$, and $U(x,y)=U(y)$ stands for the potential of applied electric field
along $y$ direction. We solve the equation $H\Psi =E\Psi $ respectively in
left barrier, middle dot and right barrier to calculate the bound state
energy within the conduction band if $V_{barrier}>V_{gate}$. According to
the semiconducting armchair boundary conditions, the quantized transverse
momentum is given as
\begin{equation}
q_{n}=\frac{(n+1/3)\pi }{W},n=0,\pm 1,\cdots
\end{equation}%
where $W$ is the width of the ribbon. Additionally, after matching the wave
function at $y=0$ and $y=L$ ($L$ is the length of dot), we can obtain
discrete energy levels because of the electrostatical confinement in the
longitudinal direction. The results are systematically shown in Fig. 1. The
bound state energy levels only exist when $E$ satisfies
\begin{equation}
E-eV_{gate}\geq \hbar v_{F}q_{n}\geq |E-eV_{barrier}|.
\end{equation}

\begin{figure}[tbp]
\centering
\subfigure[]{\label{1a}\includegraphics[width=0.4\columnwidth]{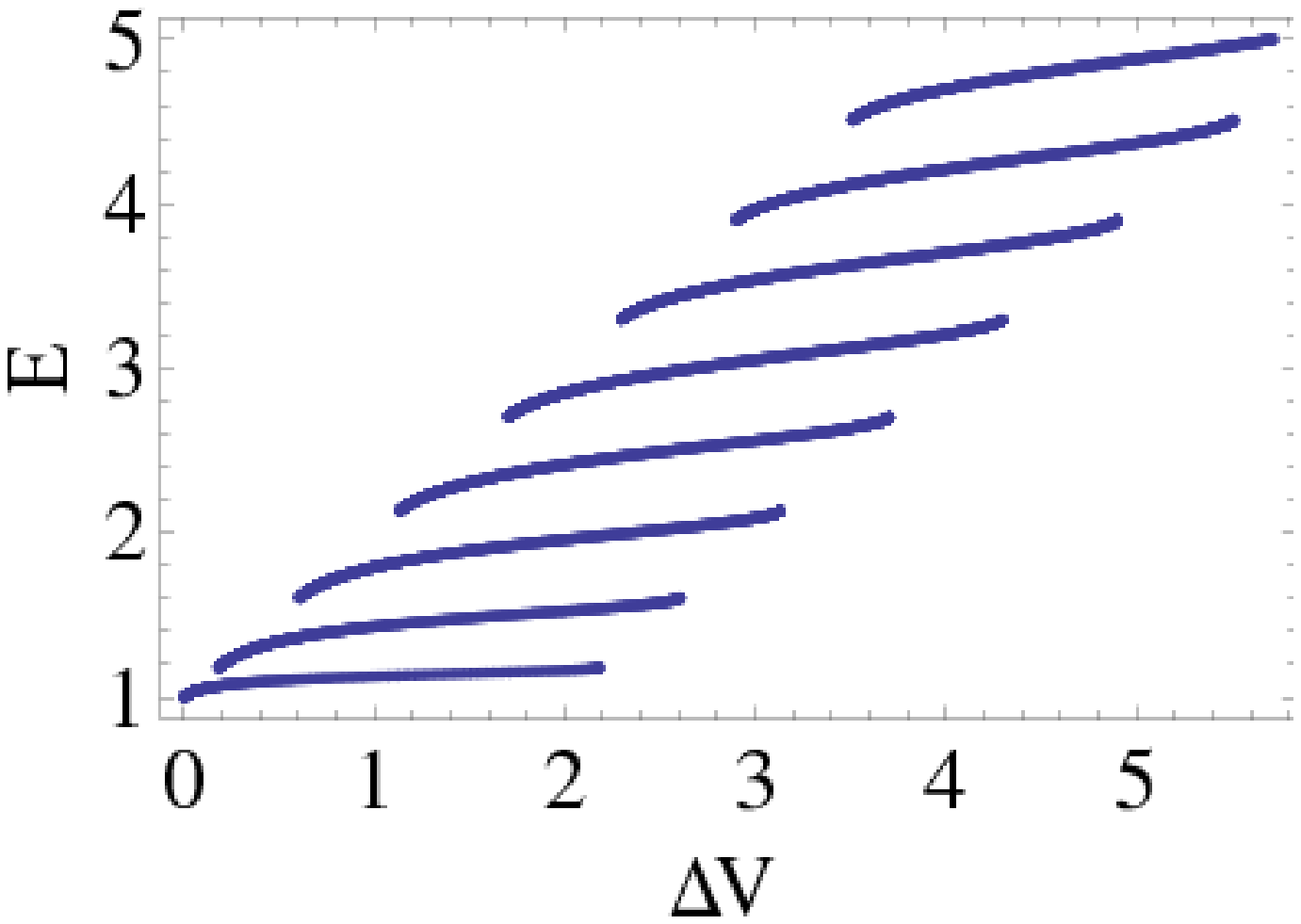}}
\subfigure[]{\label{1b}\includegraphics[width=0.4\columnwidth]{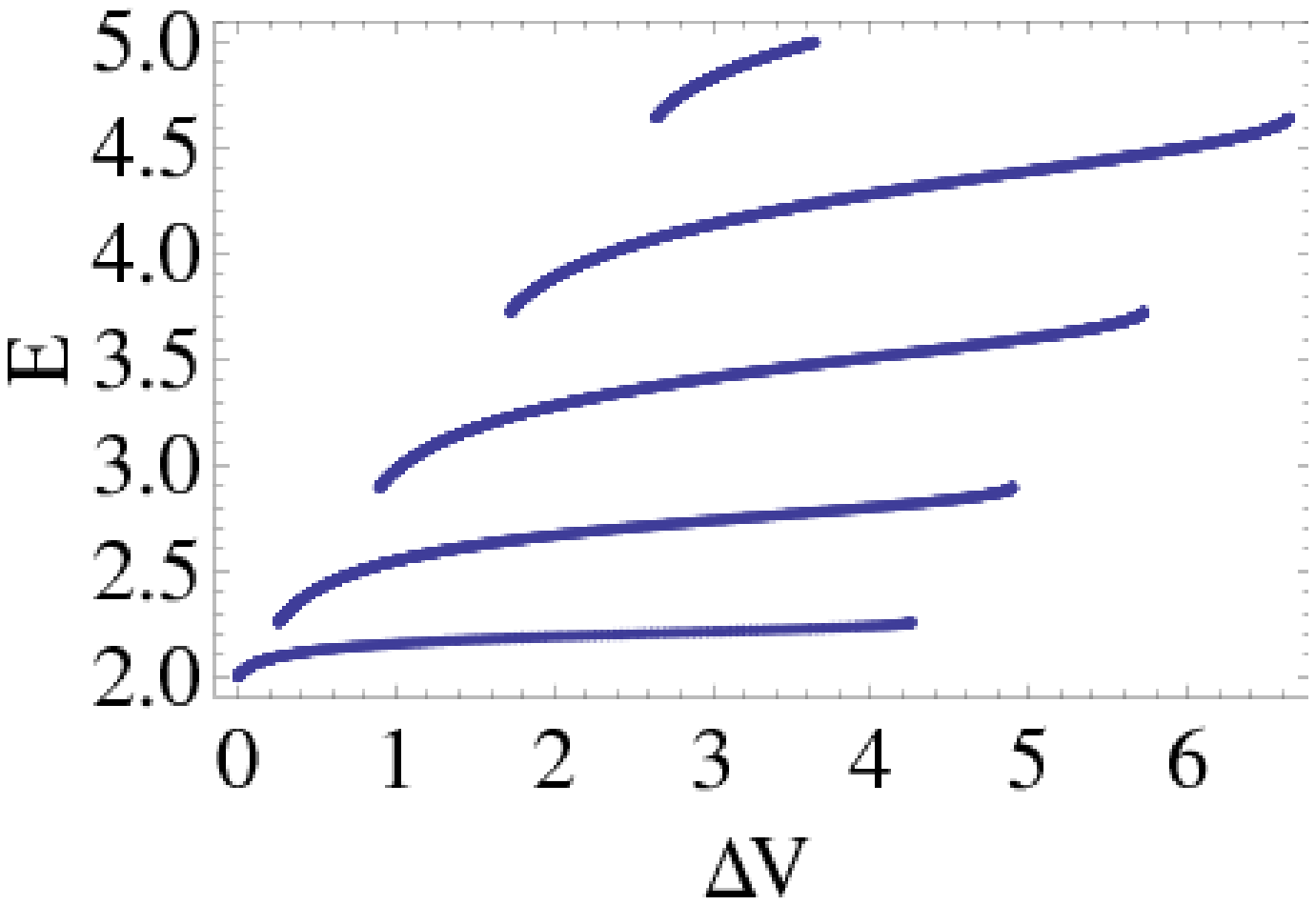}}
\caption{ Calculated energy levels in the graphene quantum dot for
different quantized transverse momentum (a) $q_{0}$, (b) $q_{-1}$.
We consider a relatively long dot $q_{0}L=5$ as in Ref.
\protect\cite{Burkard2007}. The horizontal axis stands for $\Delta
V=V_{barrier}-V_{gate}$ and the vertical axis represents the
electron bound energy, both in a unit of $\hbar v_{F}q_{0}$. }
\end{figure}

\textit{Coulomb blockade behavior. }In the following analysis, we consider
the five lowest quantized energy levels in the graphene quantum dot at
potential barrier $\Delta V=1$ in a unit of $\hbar v_{F}q_{0}$, of which
three belong to $q_{0}$, and two belong to $q_{-1}$. Here we set the
nanoribbon width $W=45$ nm according to Ref. \cite{Ensslin2009-2}, then $%
\hbar v_{F}q_{0}=0.015$ eV. First, we adopt the Constant Interaction (CI)
model \cite{Beenakker1990}, which assumes that the Coulomb interaction
between the electrons is independent of the number $N$ of electrons in the
dot and can be described by a constant capacitance $C$, and estimate the
charging energy $e^{2}/C\approx 1$ in a unit of $\hbar v_{F}q_{0}$, which is
constant for a given graphene quantum dot geometry and in the order of
experimental measurements \cite{Geim2008,Ensslin2009-2}. In this model, the
additional energy is $e^{2}/C+\triangle E$, where $\triangle E$ is the
energy difference between two consecutive states. The CB peaks will appear
if the gate charging can supplement the additional energy for next electron.
Then we use the method described in Ref. \cite{Beenakker1990} to discuss in
detail the single electron tunneling phenomenon at low temperature in the
weak tunneling regime ($h\Gamma \ll k_{B}T\ll \Delta E$). The linear
response conductance is
\begin{equation}
G=-\frac{e^{2}}{2k_{B}T}\underset{N}{\sum }\Gamma _{N}\ f^{\prime
}(E_{N}+U(N)-U(N-1)-E_{F}),
\end{equation}%
where $f(x)=\frac{1}{1+e^{\frac{x}{k_{B}T}}}$ is the Fermi-Dirac
distribution function, $E_{N}$ is the energy of the top filled single
electron state for a $N$ electron dot, $U(N)=(Ne^{2})/2C-Ne\phi _{ext}$ in
which $\phi _{ext}=\phi _{0}+\alpha V_{gate}$, and $E_{F}$ is the chemical
potential of the lead. The width of localized energy $\Gamma $ in the above
equation is determined by tunneling though the classically forbidden region
\cite{Chakraborty2007}
\begin{equation}
T=exp\left( -\int_{L}^{L+d}\left\vert k^{^{\prime }}\right\vert dy\right) ,
\end{equation}%
where $k^{\prime }=i\sqrt{q_{n}^{2}-((E-eV_{barrier})/\hbar v_{F})^{2}}$ is
the vanishing wave vector in the barrier region and $d$ is the width of each
barrier (setting $q_{0}d=1$). Notice that the tunneling rate will be
suppressed at a large $k^{\prime }$.

\begin{figure}[tbp]
\includegraphics[width=0.6\columnwidth]{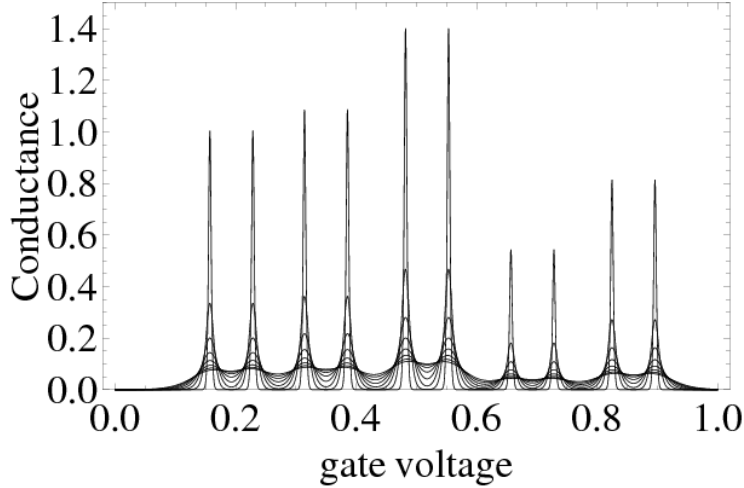}
\caption{ Coulomb blockade oscillations for a graphene quantum dot at
different temperatures. From top to bottom, $%
k_{B}T=0.02,0.06,0.10,0.14,0.18,0.22,0.26,0.30$, in a unit of $\hbar
v_{F}q_{0}$. Spin freedom has been considered. The gate voltage is in a unit
of eV and all the conductance peak heights are normalized by the first peak
at $k_{B}T=0.02$.}
\end{figure}

Fig. 2 shows the calculated conductance of graphene quantum dot as a
function of gate voltage at different temperatures, in which we take
consideration of spin freedom. We find that the derived function of
Fermi-Dirac distribution performs as a $Delta$ function and it induces a
conductance peak at position $E_{N}+U(N)-U(N-1)=E_{F}$. The peak height
maximum $G_{max}$, according to Eq. (5), is given by $G_{max}=\frac{e^{2}}{%
8k_{B}T}\Gamma $, which decreases linearly with increasing temperature in
the quantum CB regime. Also, the $N$th peak probes the specific excitation
spectrum around $E_{N}$, the quantum CB regime therefore usually shows
randomly varying peak heights \cite%
{Geim2008,Brouwer2009,Wurm2009,Libisch2009}. We would also like to mention
that in recent experiments \cite{Geim2008,Ensslin2008-1,Ensslin2008-2}
graphene dots have most likely rough edges which can induce electron
scattering at the atomically sharp edges. Nevertheless we can find the lack
of rough edges effect in the model described here does not change
qualitatively the main features of the CB. Furthermore, the line shape of
conductance can be obtained as
\begin{equation}
\frac{G}{G_{max}}=\text{cosh}^{-2}(\frac{\delta }{2k_{B}T}),\text{ }\delta
=E_{N}+U(N)-U(N-1)-E_{F},
\end{equation}%
which is consistent with the recent experimental observations in single
electron transistor on graphene nanoribbon \cite{Ensslin2009-2}.
\begin{figure}[tbp]
\centering
\subfigure[]{\label{3a}\includegraphics[width=0.5\columnwidth]{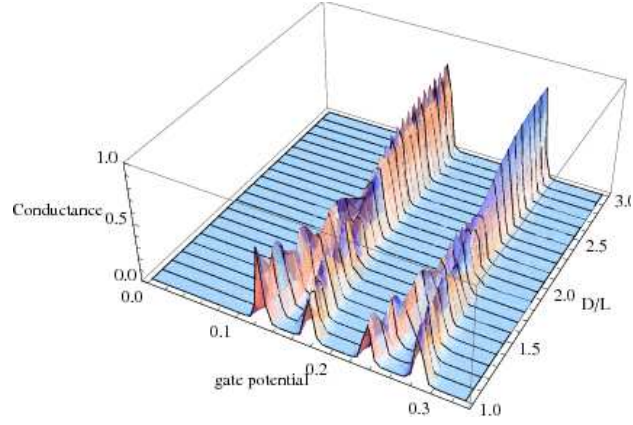}} %
\subfigure[]{\label{3b}\includegraphics[width=0.5\columnwidth]{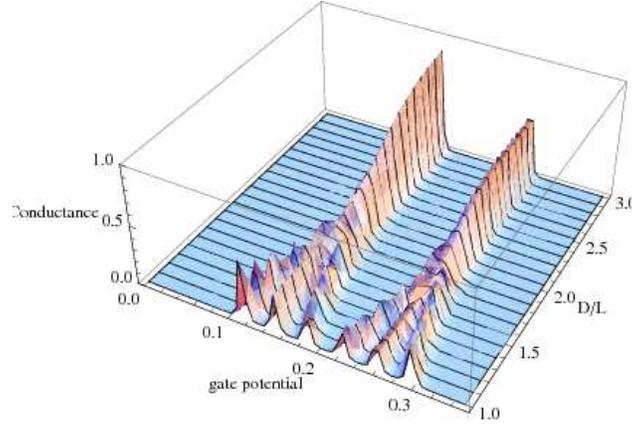}} %
\subfigure[]{\label{3c}\includegraphics[width=0.5\columnwidth]{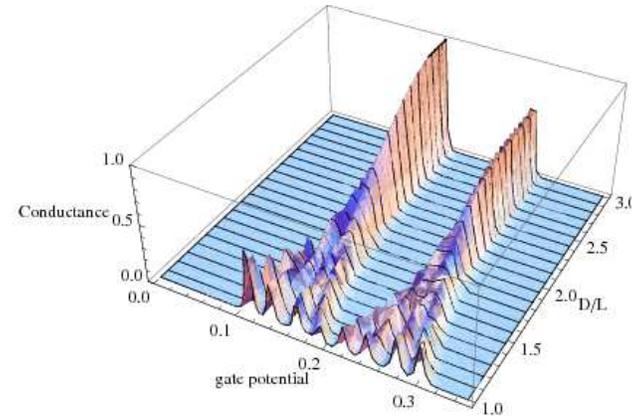}}
\caption{ Plots showing the evolution of the Coulomb blockade
oscillations as a function of the interdot tunneling strength $D/L$.
$D$ is the distance between two nearest graphene dots and $L$ is the
length of individual dot. Here we demonstrate the conductance
spectrum of multiple coupled graphene dots for the case of (a) two,
(b) three, (c) four. The gate voltage is in a unit of eV and the
conductance is normalized by the highest peak. }
\end{figure}

\textit{Graphene quantum dot array. }In the following section, we
investigate the conductance spectrum multiple coupled graphene quantum dot
systems realized by assembling the above single dot separated from each
other by a metallic gate. The type of coupling between quantum dots
determines the character of the electronic states and the nature of
transport through the quantum dot array. Here we use a Hubbard model to
account for the effects of nearest interdot tunneling which is much smaller
than the Coulomb energy of each dot, and neglect the long-ranged Coulomb
interactions which are screened by the metallic gate \cite{Burkard2007}.

We consider a doubly spin-degenerate state in each graphene quantum dot and
to study the tunneling phenomenon in coupled dots. Due to the weak
spin-orbit coupling in graphene, the spin degeneracy cannot be easily broken
if without applying magnetic field. The tunneling strength in this case is
determined by the interdot distance exponentially \cite{Burkard2007}. If the
distance between the nearest dots is quite large compared with the dot size,
the tunneling strength approaches zero, and the individual dot energy levels
are conserved. As the interdot distance is decreased, the tunneling between
nearest dots cannot be ignored, and the energy states are split into
subbands, while spin degeneracy is still remained. As the discrete energy
levels including the intradot Coulomb repulsion have been calculated above,
it is convenient to get the tunneling coupled energy levels via direct
diagonalization \cite{Dagotto1991}.

Fig. 3 shows the calculated conductance of two, three and four dots, from
which, we find a remarkably phase transition for tunneling coupled graphene
quantum dot array \cite{DasSarma1994}. For weak interdot tunneling, i.e.,
when $D/L$ is large ($D$ is the distance between nearest dots and $L$ is the
length of a single dot), the Coulomb blockade of individual dots is
maintained. The number of peaks is equal to the number of energy levels
within a single graphene dot without interacting with surrounding dots. Each
peak represents the addition of electrons to the array, one to each graphene
dot at the same time. For the moment, transport through double graphene dots
in this regime has been reported \cite{Ensslin2009-1}. For intermediate $D/L$%
, the Coulomb blockade of individual dots is destroyed and a collective
Coulomb blockade phenomenon appears. The original two energy states are
split into two subbands and each capacitance peak is split into several
peaks equal to the dot number. For stronger interdot tunneling strength,
subbands will cross each other and the Coulomb blockade is destroyed
altogether. All these theoretical predictions can be verified in the near
future experiments.

\textit{Conclusion.} In summary, we investigate the quantum dot behavior of
electrostatically confined graphene nanoribbon. We study the single electron
tunneling phenomenon at low temperatures of this system. The periodicity,
amplitude and line shape of the CB oscillations are discussed in detail.
Also, we have presented calculations of conductance spectra for multiple
coupled graphene dots and find a phase transition from individual CB to
collective CB as the coupling strength increases. The results presented here
are important to provide necessary information for future experimental work.

\textit{Acknowledgement.} This work at USTC was funded by National Basic
Research Programme of China (Grants No. 2006CB921900 and No. 2009CB929600),
the Innovation funds from Chinese Academy of Sciences, and National Natural
Science Foundation of China (Grants No. 10604052 and No. 10874163 and
No.10804104).



\begin{thebibliography}{99}
\bibitem{Geim2007} A. K. Geim and K. S. Novoselov, Nature Materials \textbf{%
6,} 183 (2008).

\bibitem{Geim2009} A. H. Castro Neto et al., Rev. Mod. Phys. \textbf{81,}
109 (2009).

\bibitem{Efetov2007} P. G. Silvestrov and K. B. Efetov, Phys. Rev. Lett.
\textbf{98}, 016802 (2007).

\bibitem{Burkard2007} B. Trauzettel, D. Bulaev, D. Loss and G. Burkard,
Nature Physics \textbf{3,} 192 (2007).

\bibitem{Peeter2007} J. Milton Pereira, P. Vasilopoulos, and F. M. Peeters,
Nano Lett. \textbf{7,} 946 (2007).

\bibitem{Egger2007} A. De Martino, L. Dell'Anna, and R. Egger, Phys. Rev.
Lett. \textbf{98,} 066802 (2007).

\bibitem{Geim2008} L. Pononmarenko et al., Science \textbf{320}, 356 (2008).

\bibitem{Ensslin2008-1} C. Stampfer et al., Appl. Phys. Lett. \textbf{92},
012102 (2008).

\bibitem{Ensslin2008-2} C. Stampfer et al., Nano Lett. \textbf{8}, 2378
(2008).

\bibitem{Ensslin2009-1} F. Molitor et al., arXiv:0905.0660.

\bibitem{Brouwer2009} J. H. Bardarson, M. Titov, and P. W. Brouwer, Phys.
Rev. Lett. \textbf{102,} 226803 (2009).

\bibitem{Wurm2009} J. Wurm et al., Phys. Rev. Lett. \textbf{102}, 056806
(2009).

\bibitem{Beenakker1990} C. W. J. Bennakker, Phys. Rev. B \textbf{44}, 1646
(1990).

\bibitem{Chakraborty2007} H.-Y. Chen, V. Apalkov, and T. Chakraborty, Phys.
Rev. Lett. \textbf{98}, 186803 (2007).

\bibitem{Libisch2009} F. Libisch, C. Stampfer, and J. Burgdorfer, Phys. Rev.
B \textbf{79}, 115423 (2009).

\bibitem{Ensslin2009-2} C. Stampfer et al., Phys. Rev. Lett. \textbf{102},
056403 (2009).

\bibitem{DasSarma1994} C. A. Stafford and S. Das Sarma, Phys. Rev. Lett.
\textbf{72}, 3590 (1994).

\bibitem{Dagotto1991} E. Dagotto et al., Phys. Rev. Lett. \textbf{67}, 1918
(1991).
\end{thebibliography}
\end{document}